\documentclass[3p,twocolumn]{elsarticle}

\usepackage{subcaption} 
\usepackage{graphicx} 
\usepackage{circuitikz} 
\usepackage{float} 

\tikzset{>=latex} 
\usepackage{amsmath} 
\usepackage{amssymb}
\usepackage{cuted}
\usepackage{changepage}
\usepackage[linesnumbered,ruled]{algorithm2e}

\usepackage{graphicx}

\newcommand{\simm}{\raise.17ex\hbox{$\scriptstyle\sim$}}
\newcommand{\bmat}{\begin{bmatrix}}
\newcommand{\emat}{\end{bmatrix}}

\begin{document}
\begin{frontmatter}

\title{Humidity-Aware Model Predictive Control for Residential Air Conditioning: A Field Study}

\author[purdue]{Elias N. Pergantis}
\author[purdue]{Parveen Dhillon}
\author[purdue]{Levi D. Reyes Premer}
\author[purdue]{Alex H. Lee}
\author[purdue]{Davide Ziviani}
\author[purdue]{Kevin J. Kircher \corref{correspondent}}

\address[purdue]{Center for High Performance Buildings, Purdue University, 177 S Russell St, West Lafayette, IN 47907, USA}
\cortext[correspondent]{Corresponding author: \texttt{kircher@purdue.edu}}

\begin{abstract}
Model predictive control of residential air conditioning could reduce energy costs and greenhouse gas emissions while maintaining or improving occupants' thermal comfort. However, most approaches to predictive air conditioning control either do not model indoor humidity or treat it as constant. This simplification stems from challenges with modeling indoor humidity dynamics, particularly the high-order, nonlinear equations that govern heat and mass transfer between the air conditioner's evaporator coil and the indoor air. This paper develops a machine-learning approach to modeling indoor humidity dynamics that is suitable for real-world deployment at scale. This study then investigates the value of humidity modeling in four field tests of predictive control in an occupied house. The four field tests evaluate two different building models: One with constant humidity and one with time-varying humidity. Each modeling approach is tested in two different predictive controllers: One that focuses on reducing energy costs and one that focuses on constraining electric power below a utility-specified threshold. The two models lead to similar performance for reducing energy costs. Combining the results of this study and a prior heating study of the same house, the estimated year-round energy cost savings were \$340--497 or 22--31\% (95\% confidence intervals); these savings were consistent across both humidity models. However, in the demand response tests, the simplifying assumption of constant humidity led to far more frequent and severe violations of the power constraint.
\end{abstract}


\end{frontmatter}

\section{Introduction}

\subsection{Supervisory air conditioning control}

Air conditioning is a potentially life-saving service that will be increasingly vital in a warming world. However, air conditioning uses one-tenth of global electricity today, and experts expect global demand for air conditioning to triple by 2050 \cite{iea2018future}. This growth in electricity demand could increase energy costs, air pollution, and greenhouse gas emissions, and could require costly build-out of electrical infrastructure or increase the frequency of blackouts \cite{horowitz2019distribution,sharma2021major,priyadarshan2024edgie}. Improving the energy efficiency of air conditioning would mitigate all of these risks.

\begin{figure*}
\centering
\includegraphics[width=0.8\textwidth]{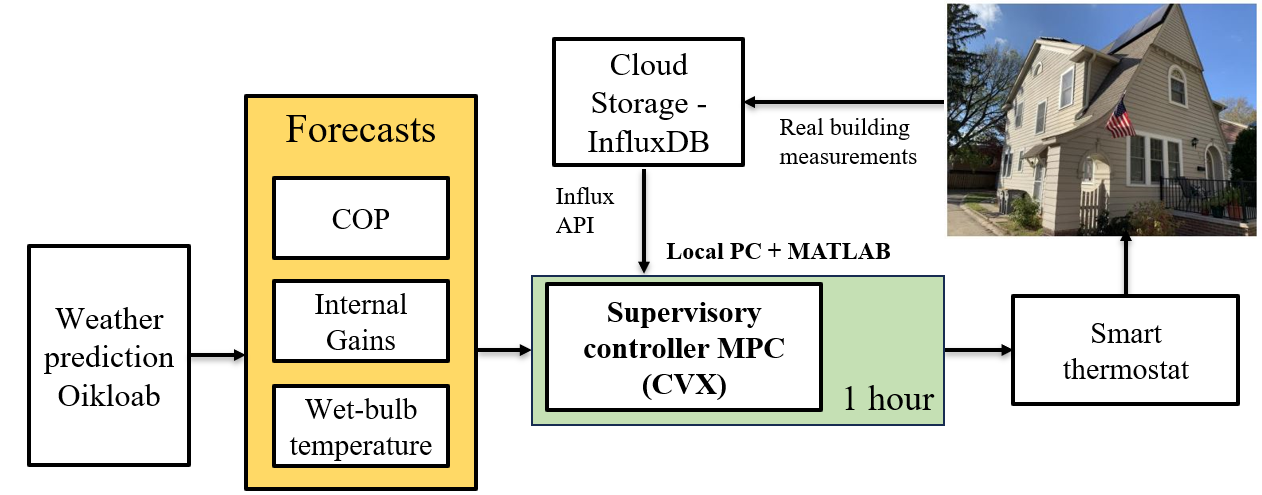}
\caption{Flow-chart of the supervisory control system. Weather information (temperature, humidity, solar irradiance, wind speed) is obtained from a weather service and sent to the controller alongside disturbance predictions and real-time building measurements. The selected indoor air temperature set-point is sent to the smart thermostat and implemented in the house.}
\label{summer_flowchart}
\end{figure*}

Supervisory control of heating, ventilation, and air conditioning (HVAC) equipment is one way to improve the energy efficiency of air conditioning \cite{drgovna2020all,Kapsalaki}. Supervisory HVAC control systems dynamically adjust set-points -- such as indoor air temperatures, compressor speeds, or fan speeds -- that device-level control systems track. One prevalent supervisory control methodology is model predictive control (MPC) \cite{bunning2020experimental,KIM201849}. This approach, illustrated in Fig. \ref{summer_flowchart}, typically uses models of the HVAC equipment and building, as well as forecasts of weather and occupancy, to repeatedly decide control actions that optimize performance objectives over a receding prediction horizon \cite{drgovna2020all}.

Air conditioners affect, and are affected by, both the temperature and humidity of the indoor air. The extent to which humidity modeling influences controller performance is an open question. The answer likely depends on the equipment configuration, building type, climate zone, and control objectives. Pergantis et al. \cite{pergantis2024field} recently reviewed the literature on field demonstrations of supervisory HVAC control in residential buildings and found that no air conditioning study accounted for time-varying indoor humidity and dehumidification load. However, in commercial buildings, other researchers have shown in numerous simulation studies \cite{RAMAN2020115765, MEI2017439}, as well as three experimental studies \cite{YANGExperiment, YANGExperiment2, YANG2020115147}, that neglecting humidity can decrease comfort and controller performance. These studies modeled the indoor humidity as a separate state in a hygrothermal model, impacted by the indoor dry-bulb temperature and HVAC dehumidification. However, there is significant engineering effort in training the multiple correlations to model the indoor heat exchanger \cite{RAMAN2020115765, RAMANmanyclimates}, as well as in solving the resulting nonconvex optimization problem \cite{RAMAN2020115765}. Several review papers \cite{killian2016ten, YAO2021107952, AFRAM2014343} have noted the challenges related to modeling dehumidification in HVAC systems, as well as the lack of humidity modeling in many simulation studies \cite{HU2014233, MAYOSTENDORP2011428, LIANG2015256}. For these reasons, most experimental demonstrations of supervisory air conditioning control model only temperature dynamics and make the simplifying assumption that humidity is constant \cite{HILLIARD2017326, HUA2024129883, WEST2014271, HUANG2015203}, neglecting humidity dynamics driven by time-varying occupant behavior.

This paper presents the first experimental evaluation of an MPC-based supervisory controller in a residential building that considers dehumidification in the problem formulation. A simplified approach is developed to reduce the complexity of training a humidity model and modeling the indoor heat exchanger. In this approach, the indoor humidity state, defined here as the return air (mixed from the whole building) wet-bulb temperature, is predicted through machine learning. This approach is well-suited for residential buildings, where active humidity control is typically not required \cite{TenWolde2008}, but where better predicting the impact of humidity on energy use can improve controller performance. Two methods are formulated to predict an air conditioner's total electricity demand, which can be decomposed into latent (humidity-change) and sensible (temperature-change) components. The first method generates load predictions wherein the sensible heat ratio (SHR, the ratio of sensible load to total load) varies over time. The second method, by contrast, assumes a constant SHR. Following the naming convention of  Raman et al. \cite{RAMAN2020115765, RAMANmanyclimates}, we refer to the first load prediction method as the ``latent'' model and the second as the ``sensible'' model. Each load prediction model is tested in two MPC implementations, resulting in a total of four experiments. The first MPC implementation aims to minimize energy cost and maximize thermal comfort. The second MPC implementation also aims to minimize energy and maximize comfort, but also seeks to constrain electrical power below a utility-specified threshold between 4 PM and 8 PM. This power constraint emulates a demand response call during the hours when residential electricity demand typically peaks. For energy cost minimization, the latent and sensible models give similar performance, which corroborates the findings of previous authors \cite{blum2019practical,wang2023field} on the impact of model accuracy on energy savings. For power limiting, although both latent and sensible models enable some load shifting, the sensible model leads to significantly more frequent, longer, and more severe constraint violations.

\subsection{Contributions} 

This paper makes five main contributions.
\begin{enumerate}

\item A summary of the state of the art of supervisory control experiments that consider humidity in residential or commercial buildings. This represents the first comprehensive synthesis of this scientific literature. Further, advances in simulation studies since the last review \cite{RAMAN2020115765} are presented to highlight trends in the field. 

\item Year-round performance of MPC in a real house. As far as the authors are aware, no supervisory control experiment in a single residential building has reported both heating and cooling performance. Using 38 days of summer testing from this study and 33 days of winter testing from \cite{pergantis2024field}, it is estimated that the controller reduced the annual cooling and heating energy cost by \$419 (95\% confidence interval: \$340 to 497), or 27\% (22 to 32\%). This represents a significant data point highlighting the reliability and performance of predictive control for residential HVAC.

\item Low-effort consideration of humidity. A novel physics-inspired machine learning method is presented for predicting the real-time indoor wet-bulb temperature. This method improves the accuracy of the electrical power prediction and eliminates the need for detailed simulations of the evaporator, a major bottleneck in humidity-aware predictive control \cite{RAMAN2020115765, RAMANmanyclimates}. This method is used in conjunction with a Predicted Percentage Dissatisfied (PPD) model for occupants' comfort. Comfort surveys indicated a single response instance of mild temperature discomfort, while analysis of the humidity and temperature time series \cite{enescu2017review,tartarini2020pythermalcomfort} indicated a total of two hours of mild discomfort in over 38 days of active control. 

\item The first experimental demonstration of power-limiting control for a heat pump in a real home. The controller reduced peak electricity demand from air conditioning by 88\% in the 4 to 8 PM demand response window while maintaining occupant comfort.

\item Discussion of deployment challenges. Although many studies discuss the benefits of supervisory HVAC control, few discuss the practical challenges that arise during deployment. Other studies have noted the need for discussion of deployment challenges \cite{blum2022field}. This paper presents a detailed summary of the major obstacles faced in model development and experimental deployment of the controller, as well as the solutions found. 

\end{enumerate}

\subsection{Organization of this study} 
Section \ref{reviewSection} of this paper reviews the state-of-the-art experimental and simulation work on supervisory HVAC control with a focus on humidity considerations. Section \ref{methodsSection} discusses the building envelope modeling, learning, and control methods underlying the supervisory control system. Section \ref{systemSection} discusses the test site and instrumentation setup. Section \ref{resultsSection} presents field results. Section \ref{discussionSection} discusses practical considerations for implementing this system and supervisory controllers more generally in homes with central air conditioning.

\section{Background and state of the art}
\label{reviewSection}

This section presents a review of the treatment of indoor humidity in supervisory HVAC control. The purpose of this review is to identify the latest numerical and experimental investigations on supervisory control that account for humidity and the challenges involved. The literature reviews of Blum et al. \cite{blum2022field} and Pergantis et al. \cite{pergantis2024field} found that only three studies out of the total 25 supervisory control experiments actively considered latent cooling loads, while no residential study included latent loads in the optimization formulation. This section therefore focuses on two questions: (1) Why have so few studies considered humidity in the problem formulation? (2) What research gaps remain in control-oriented humidity modeling?

\subsection{Simulations}

\textit{Hygrothermal models.}
In the majority of the reviewed studies, a hygrothermal model is adopted to couple temperature and humidity responses in a building. The term ``hygrothermal'' refers to models that include coupling between the air moisture dynamics and air temperature dynamics. Key features of hygrothermal models are (1) humidity dynamics in the building, and (2) the cooling heat exchanger model. The humidity dynamics are usually presented in the form of a mass conservation ordinary differential equation that includes sources (human activity), sinks (dehumidification), transfer between zones and the ambient environment (resistances through walls, buffers), and storage \cite{KUNZELoriginal}. The source term is typically a function of the number of people \cite{RAMAN2020115765,  MEI2017439, GOYAL2012332, Cai, WANG2013999}. The physical parameters (resistance through the walls, storage capacity) can either be identified based on real building data \cite{Cai, YANG2019106326} or trained using physics-based simulators, such as EnergyPlus \cite{RAMAN2020115765, beam}. Other work has also derived physical equations for a particular prototype room \cite{YANG201825} using location-specific building geometries and physical parameters.

\textit{Heat exchanger modeling.} One of the major challenges of modeling humidity dynamics in buildings arises from the complex dehumidification phenomena occurring across the indoor heat exchanger. One class of models involves solving systems of partial differential equations based on the geometry of the heat exchanger and its dry, wet, and mixed regions \cite{Zhoubraun}.  The structure and parameters of the governing equations vary between types of heat exchangers, such as air-to-water or air-to-refrigerant \cite{Zhoubraun}.

Since time constants are typically on the order of hours for building thermal dynamics and on the order of minutes for the heat exchanger \cite{RAMAN2020115765, Zhoubraun}, static heat exchanger models suffice. To extend the capabilities of static models as the inlet conditions change, Raman et al. \cite{RAMAN2020115765} used 1,159 temperature and humidity binds to characterize the inlet conditions to the heat exchanger, with each bin using a 5th-degree polynomial model. The resulting root-mean-square error (RMSE) for the prediction of the supply air temperature and relative humidity were 0.28 $^\circ$C and 1\%, respectively. The model was trained using EnergyPlus \cite{BCVTB}. Other studies have similarly constructed multiple regression models based on the temperature and humidity operating regions \cite{SCHWINGSHACKL2016250} or have simplified governing equations into large sparse systems of linear ODEs \cite{DULLINGER20181646}.

\textit{Heat exchanger model-free studies.} It is possible to incorporate a time-varying humidity term into a gray-box building model (meaning a model that has some physics-based components and some data-driven components), and to couple it with either fixed heat exchanger outlet conditions \cite{YANGExperiment,YANGExperiment2}, or multiple bins as per Raman et al. \cite{RAMAN2020115765} to obtain a linear humidity model. However, other researchers have developed model-free approaches to predict dehumidification effects without modeling the heat exchanger outlet conditions \cite{XI2007897, JIANG2020114174, XIAOmodelfree}. In \cite{XIAOmodelfree}, a long-short-term memory neural network and a recurrent neural network were used in tandem to predict future thermal comfort (specifically, the Predicted Mean Vote) as well as the temperature and humidity in a multi-zone commercial building. The resulting nonlinear MPC problem was solved using an interior point method with explicit constraints. The training period was 14 days, using an 80/20\% split between training and validation data. This is significantly shorter than other studies ({\it e.g.}, \cite{bunning2022physics}) due to the shorter time step of 10 minutes. In \cite{JIANG2020114174}, an Adaptive Neuro-Fuzzy Inference System was developed to dynamically predict a liquid desiccant air conditioning system's outlet air temperature and humidity ratio under varied input conditions. The study did not consider the system energy consumption model and solved the optimization problem using a genetic algorithm. In \cite{XI2007897}, support vector regression was used to identify future dynamics for temperature and humidity given previous actions. Experimental results showed good control performance in terms of reference tracking and steady-state errors while outperforming a baseline neural fuzzy controller.

The above studies all yield black-box models (meaning purely data-driven models with no physics-based components)  and nonconvex optimization formulations due to coupled dynamics between the indoor air's temperature and humidity. This paper, by contrast, develops a method that predicts SHR solely from historical humidity conditions, yielding a model with no explicit humidity-temperature coupling. This is done in an open-loop fashion, following foundational work on commercial chillers \cite{braun1990reducing}. In the context of this work, an open-loop model is one that does not explicitly consider the heat exchanger's outlet conditions, or any other measured building state, to perform a mass balance. Instead, open-loop models forecast the humidity state or SHR based on future exogenous inputs such as predictions of the outdoor air temperature and humidity. While open-loop models do not use measured building states in real time, they may use historical measurements for model training. By contrast, closed-loop models use real-time measurements of building states, either in an autoregressive fashion or through a physics-based hygrothermal model.

\subsection{Experiments}
A recent literature review of supervisory control demonstrations in residential buildings \cite{pergantis2024field} found that no study has considered dehumidification or comfort due to varying indoor humidity conditions. In commercial buildings, only  three studies \cite{YANGExperiment,YANGExperiment2, YANG2020115147} have performed this in actual buildings. Yang et al. \cite{YANGExperiment}, developed an MPC formulation for active chilled beams. Linear white-box models for building energy and indoor condition predictions were used. Additionally, the Predicted Mean Vote was approximated by a linear function of temperature and humidity \cite{YANG201825}. The controller reduced energy use while improving comfort over a baseline controller; however, constant outlet temperature and humidity were assumed, which might not be appropriate in general \cite{RAMAN2020115765}, especially with variable-speed equipment. This represents a limitation of static models unless they are multi-binned. The second study \cite{YANGExperiment2} investigated control of a dedicated outdoor air system with separate sensible and latent cooling heat exchangers. The methodology was similar to the previous study, with a static heat exchanger model. Energy and comfort improvements were demonstrated over a baseline controller. In \cite{YANG2020115147}, an adaptive nonlinear machine learning approach predicted the next system states (temperature, humidity), and a linearized PPD model mapped those values into thermal comfort. In \cite{CASTILLA2014703}, a detailed PPD model was developed for a specific experimental facility.

\subsection{Research gaps}

\begin{figure}
\centering
\includegraphics[height=0.225\textheight]{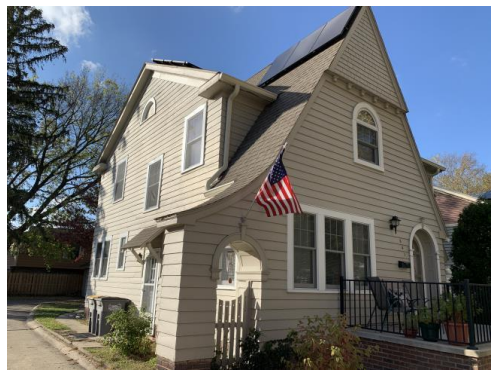}
\caption{The DC Nanogrid House is a 208 m$^2$, 1920s-era house with all-electric appliances in West Lafayette, USA.}
\label{DCHouseFig}
\end{figure}

\textit{Heat exchanger modeling complexity.}
The state-of-art supervisory control studies in buildings that consider humidity \cite{RAMAN2020115765, YANGExperiment, YANGExperiment2, YANG2020115147, RAMANmanyclimates} use static heat exchanger models and either simplify output conditions \cite{YANGExperiment, YANGExperiment2}, or model a single heat exchanger at the cost of significant development effort through high fidelity simulations \cite{RAMAN2020115765, RAMANmanyclimates}. Additionally, these models are often highly nonlinear \cite{RAMAN2020115765, YANG2020115147, XI2007897, HUrobust} and some necessitate robust or stochastic optimization techniques \cite{JIANG2020114174}.

\textit{Training model on real buildings.}
Due to the limited experimental demonstrations that consider indoor humidity variations, training accurate hygrothermal models across multiple building types using real on-site input data has not been investigated as much as for the thermal resistance-capacitance networks \cite{YANGExperiment, blum2019practical}. Currently, the majority of studies that use hydrothermal models rely on EnergyPlus to generate training data \cite{Cai}.

\textit{Lack of experiments.}
Although advanced methods that consider temperature and humidity coupling have been developed in simulation testbeds, very few studies have evaluated these methods in real buildings. Simplified humidity modeling approaches such as the ones presented in this paper could facilitate real-world implementation and testing of humidity-aware air conditioner supervisory controllers, an important step toward industry adoption.

\textit{Comfort improvements via humidity considerations.}
As corroborated by the parametric study in \cite{RAMANmanyclimates} looking at many climates, as well as the experimental works \cite{YANGExperiment, YANGExperiment2}, supervisory controls can fail to maximize their energy savings while preserving comfort. This can be seen in both cooling (experiments, simulations) and heating (simulations). However, further data points in experimental test beds are required to quantify the impact of humidity-aware control on comfort and indoor environment quality.

\begin{figure*}
\centering
\includegraphics[width=0.8\textwidth]{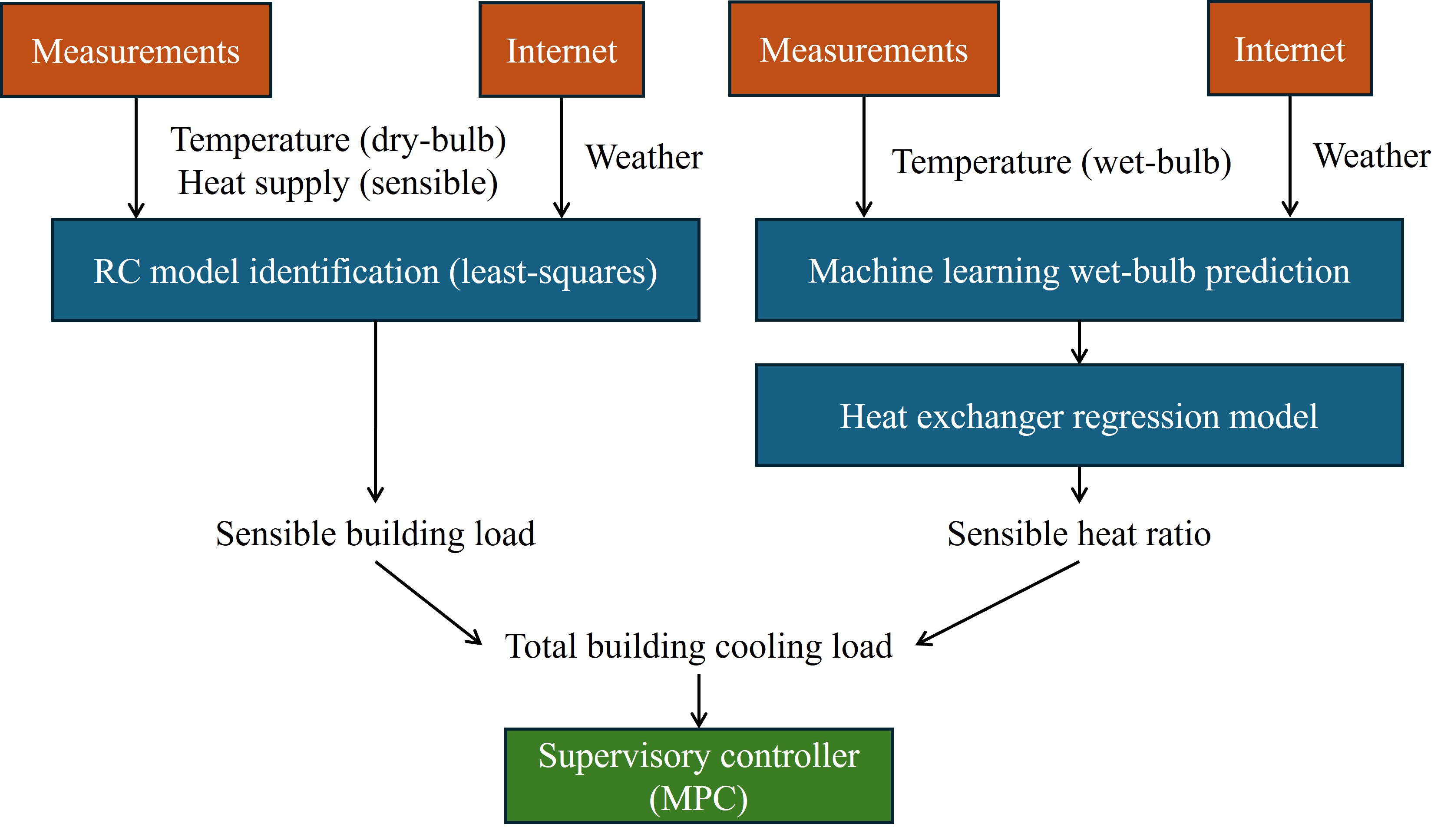}
\caption{The workflow used in this paper to model and control the building and equipment.}
\label{endotend}
\end{figure*}

\section{Modeling, learning, and control methods}
\label{methodsSection}

This section presents the methods used for modeling the system dynamics, training machine-learning models,  and controlling the HVAC equipment. Section \ref{buildingModeling} describes the process of identifying a low-order resistance-capacitance model of a detached house with a single mixed-air zone. Section \ref{equipmentModels} develops two humidity models: (1) a latent model that considers a time-varying indoor wet-bulb temperature for computing the air conditioner's coefficient and SHR, and (2) a sensible model that assumes a constant SHR and models the coefficient of performance (COP) as a function only of the outdoor temperature. Finally, Section \ref{Scheme} develops two separate supervisory controller formulations: one aimed at reducing energy costs, and one aimed at constraining power below a utility-specified threshold during a demand response window. Fig. \ref{endotend} illustrates the workflow of model development and control design presented in this section.

\subsection{Building modeling} 
\label{buildingModeling}

\begin{figure}
\centering
\begin{circuitikz}[scale=1.1, american currents] 
\ctikzset{bipoles/length=1.05cm} 
\pgfmathsetmacro{\w}{2};
\pgfmathsetmacro{\h}{1};

\node[above] at (2*\w,2*\h) {$T_m$};
\draw (2*\w,2*\h) to[battery1,*-] (2*\w,0) -- (0,0);

\node[above left] at (\w,2*\h) {$T$};
\draw (0,2*\h) to (\w,2*\h) to[R,R=$R_m$,*-] (2*\w,2*\h);

\draw (0,0) to[I,n=Qa] (0, 2*\h);
\node[right] at (Qa.s) {$\dot Q_c + \dot Q_e$};

\draw (\w,0) to[C,n=Ca] (\w,2*\h);
\node[right] at (Ca.s) {$C$};
\draw (\w,0) node[ground] {} to (\w,0);

\draw (\w,2*\h) -- (\w,3*\h) to[R,R=$R_\text{out}$,-*] (3*\w,3*\h);
\draw(3*\w,3*\h) to[battery1] (3*\w,0) -- (2*\w,0);
\node[above] at (3*\w,3*\h) {$T_{\text{out}}$};

\end{circuitikz}
\caption{A 2R1C thermal circuit model.}
\label{2R1CFig}
\end{figure}
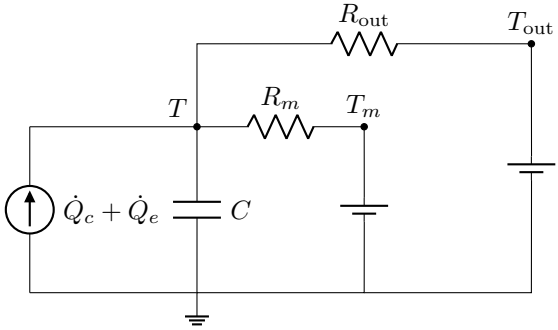

\begin{figure*}
\centering
\includegraphics[width=0.45\textwidth]{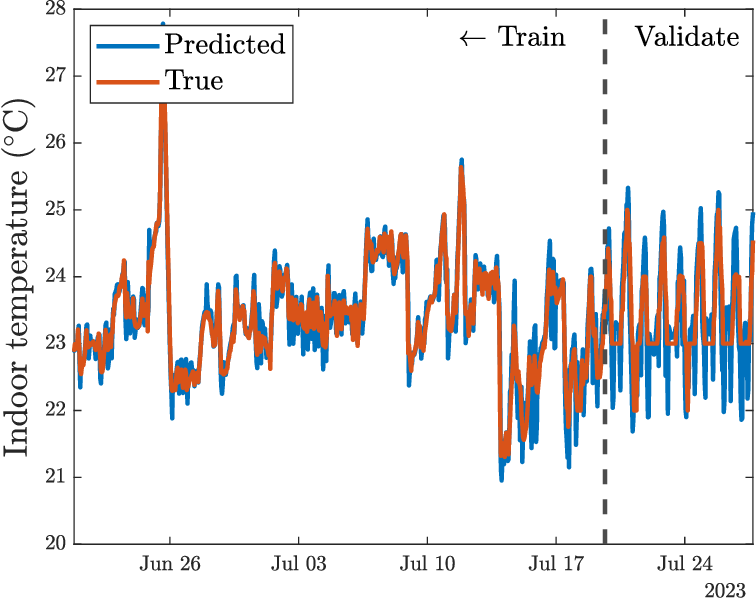}
\qquad
\includegraphics[width=0.45\textwidth]{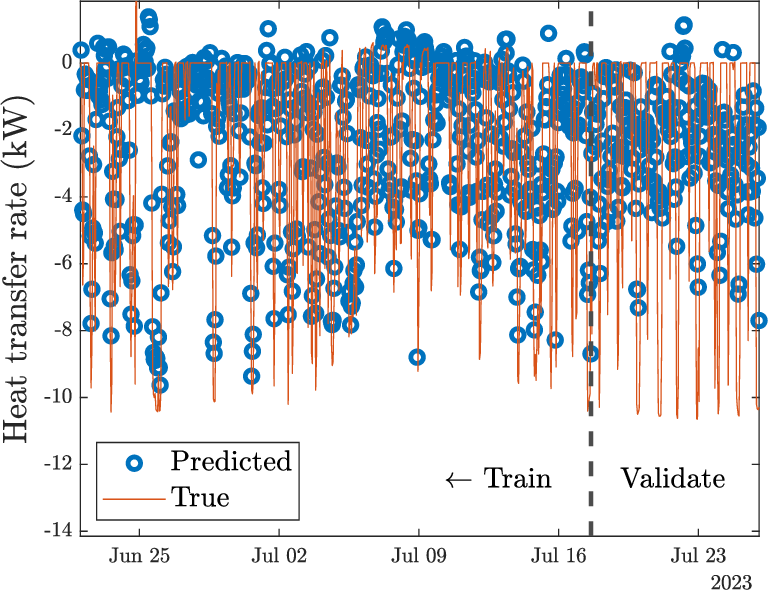}

\caption{Indoor temperature prediction errors (left) and air conditioner cooling rate predictions and measurements (right). Predictions match measurements reasonably well in both training (first month of data) and validation (last ten days).}
\label{modelFitFigure}
\end{figure*}

A linear thermal circuit model is used to capture the thermal dynamics of the test house, a typical detached single-family in Indiana, USA (Fig. \ref{DCHouseFig}). The selection of model order is governed by the need for accuracy in predicting temperatures and the required thermal building load, as well as the resulting complexity of the MPC optimization problem. In the case of central air conditioners in residential buildings, a very common system set up in North America, typically the systems can be modeled using a single mixed air zone \cite{blum2019practical}. This work uses a model with two resistances and one capacitance (2R1C), as illustrated in Fig. \ref{2R1CFig} and presented previously in \cite{pergantis2024field}. The major advantages of this model are the lack of hidden states, a simple training process via least-squares, and the ability to generate a training data set of an exogenous disturbance term that captures both model inaccuracies and heat gains from internal sources and the sun. This disturbance term can be predicted using a variety of time-series forecasting methods.

A derivation of the discrete-time governing equations and the model training process can be found in \cite{pergantis2024field}. The model was trained the model on passive observations from June 20 to July 20, 2023. These 30 days gave 720 hourly data points for each observed variable. One-quarter of the training set included days under MPC operation; the other three-quarters of the training data came from the manufacturer's default control algorith with constant indoor temperature set-points. Fig. \ref{modelFitFigure} shows the end-to-end fit, including the thermal circuit and the exogenous thermal power predictor, in training and validation data. The one-step-ahead indoor temperature predictions (left plot) match the targets with RMSE of 0.53 $^\circ$C in the validation data. The sensible cooling rate predictions matched the measurements with an RMSE of 3.6 kW. Over a steady thermostat set-point week, the RMSE for the power prediction is 0.44 kW. The model occasionally predicts heat demand on the order of 700-1000 W. While these heating predictions are unrealistic during the cooling season, the accuracy of the temperature prediction was found to be sufficient for control purposes. The observed suitability of a moderately accurate model for predictive control is consistent with prior findings, such as \cite{blum2019practical}.

\subsection{Equipment and humidity models}
\label{equipmentModels}

\textit{Heat pump COP.} During cooling operation, two physical processes occur simultaneously when return air from the house flows over the indoor heat exchanger. First, water condenses out of the air; this is latent cooling. Second, the air's dry-bulb temperature decreases; this is sensible cooling. The sensible cooling rate is equivalent to the sensible building load $\dot Q_c$, which is obtained from the thermal circuit model. The sum of the sensible and latent cooling rates is the total cooling rate, and the ratio of the sensible cooling rate to the total cooling rate is the SHR.

\begin{figure}
\centering
\includegraphics[width=0.45\textwidth]{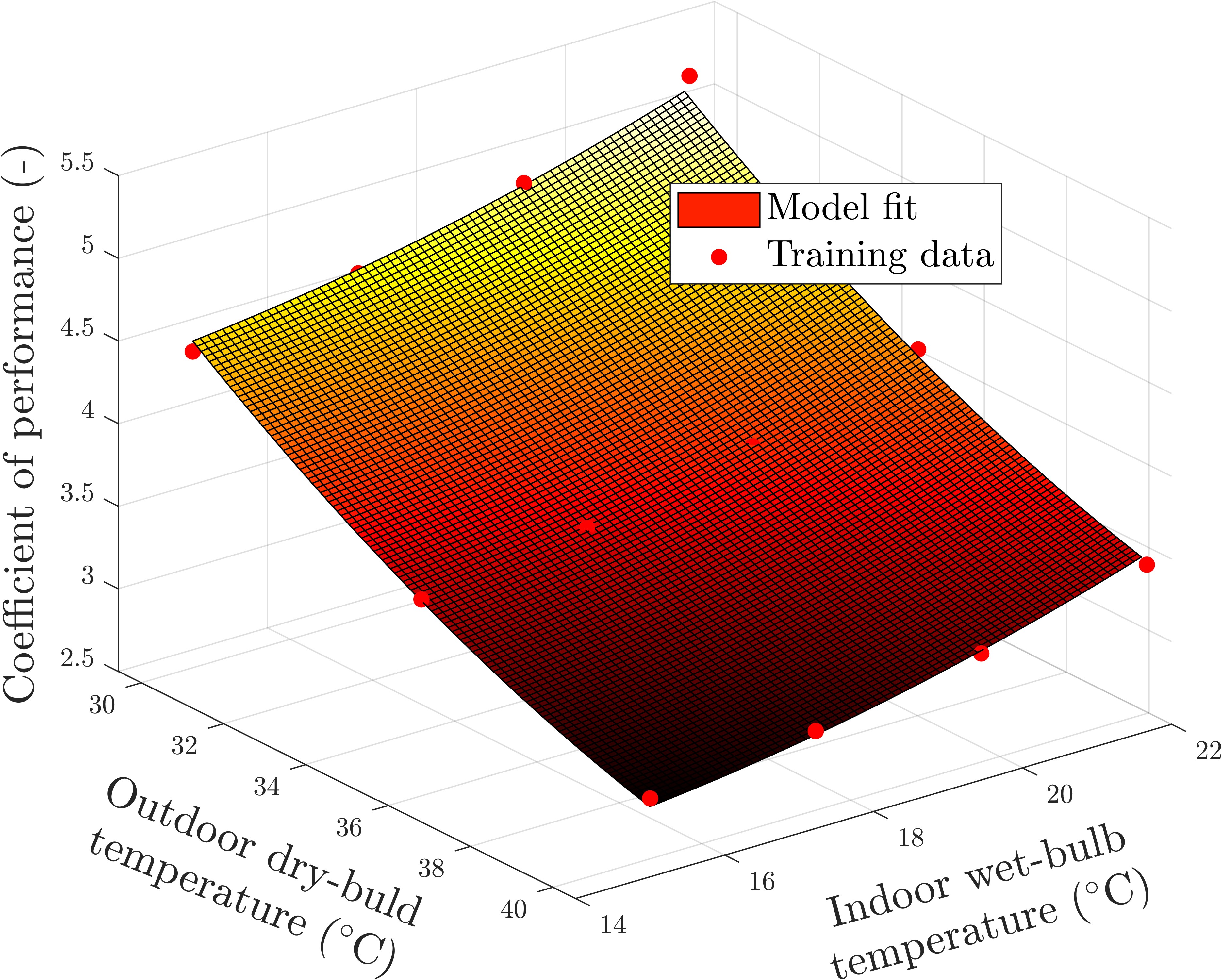}
\caption{A quadratic model of the HVAC COP as a function of indoor wet-bulb and outdoor dry-bulb temperatures fits the training data with $R^2$ = 0.99. Performance data were considered at an indoor dry-bulb temperature of 24 $^\circ$C, matching historical set-points on-site.}
\label{COPlatent}
\end{figure}

This study uses two different formulations for the heat pump COP, both based on manufacturer data. The formulations are part of a supervisory control system (MPC) that adjusts indoor temperature set-points. In the sensible formulation, the SHR is assumed constant at 0.86, consistend with the heat pump manufacturer's specification sheet. Under this condition, the COP is modeled only as a function of the outdoor air temperature. After the completion of MPC testing, the actual mean SHR was found to be 0.79 over the testing duration. The impact of this assumption is considered in the discussion section. In the latent model formulation, the SHR varies during the day. By reviewing the manufacturer's data, it was found that the SHR can be approximated as a linear function of the heat exchanger inlet air wet-bulb temperature. Additionally, the COP in the latent formulation is considered to be a function of both the outdoor air dry-bulb temperature and the indoor air wet-bulb temperature, as shown in Fig. \ref{COPlatent}.

\textit{Predicting wet-bulb temperature and SHR.} The indoor wet-bulb temperature is strongly dependent on the moisture removal by the heat pump as well as its on/off state. These effects cannot be captured in an open-loop model where there is a one-off forecast. Further, the indoor wet-bulb temperature is also a function of the outdoor air temperature, the difference between indoor and outdoor humidities that drives mass transfer through the building envelope, the time of day (which correlates with people’s activity levels), and secondary identifiers such as solar radiation and wind speed (which drives infiltration effects). To account for the impact of the MPC controller adjusting the set-points to the indoor temperature, two strategies can be employed. First, a portion of the MPC data can be included during the model training process. Second, real-time online training can be implemented, incorporating the latest wet-bulb temperature measurements from the house. For simplicity, a one-time offline training was performed in this study using MPC days for 25\% of the training data set. This approach is feasible because in scenarios where most optimization objectives are fixed, such as in this study with fixed electricity prices and no grid emissions considerations, the MPC tends to follow a fairly consistent set-point pattern based on the diurnal cycle.

\begin{figure}
\centering
\includegraphics[width=0.4\textwidth]{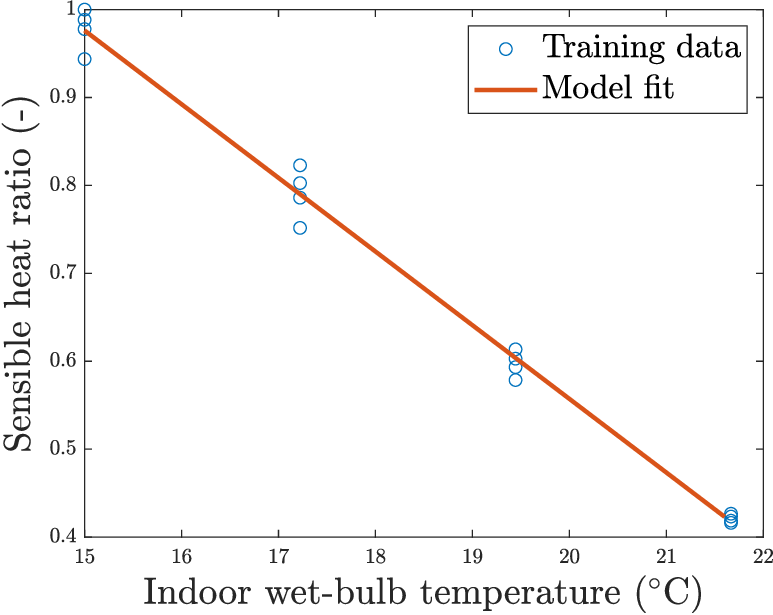}
\caption{A linear model of SHR vs. indoor wet-bulb temperature fits the training data with $R^2$ = 0.97. Variations are due to different outdoor dry-bulb temperatures.}
\label{SHRFig}
\end{figure}

\begin{figure}
\centering
\includegraphics[width=0.45\textwidth]{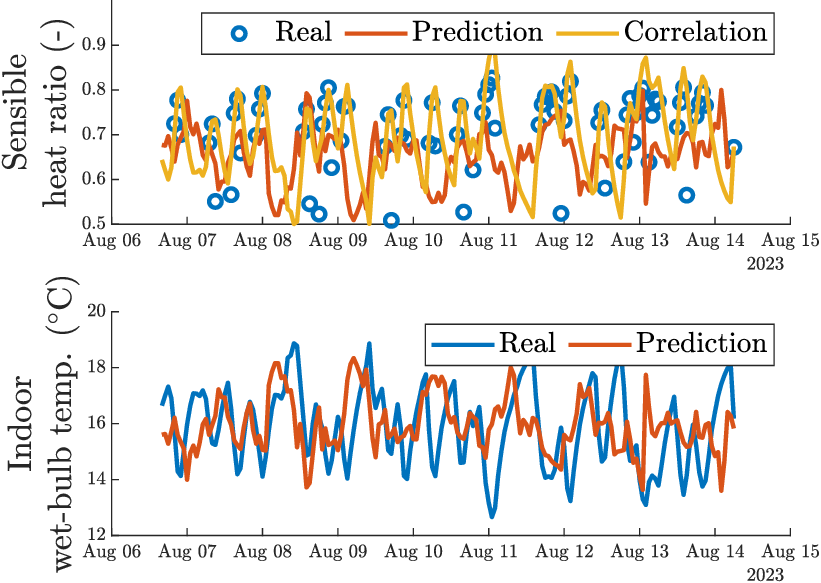}
\caption{In the field, both the simplified linear heat exchanger model (using on-site measured wet-bulb, yellow curve) and the machine-learning prediction model (top plot, orange curve) match real measurements (both plots, blue) with reasonable accuracy.}
\label{SHR_WEEK}
\end{figure}

The features used to predict to the return air wet-bulb temperature $T_{\text{wb}}$ ($^\circ$C) were the outdoor air relative humidity $RH_{\text{out}}$, the outdoor air temperature $T_{\text{out}}$ ($^\circ$C), the hour of day $h$, the solar irradiation $I_{\text{solar}}$ (kW/m$^2$), and the wind speed $w$ (m/s):
\begin{equation}
T_{\text{wb}} = \text{GPR}(RH_{\text{out}}, T_{out}, h,I_{\text{solar}}, w) .
\label{GPR}
\end{equation}
After evaluating various model structures, a Gaussian Process Regression (GPR) model performed the best in predicting the indoor wet-bulb temperature, given the same four-week training period as the thermal circuit model. The validation RMSE was 1.5 $^\circ$C in an MPC week and 0.4 $^\circ$C in a week where the thermostat set-point was constant. As Fig. \ref{SHRFig} shows, the SHR was modeled as a linear function of the inlet wet-bulb temperature, 
\begin{equation}
\text{SHR} = a  T_{\text{wb}} + b, \label{SHR}
\end{equation}
with the constants $a$ and $b$ fit using linear regression. The training data provided by the manufacturer included the sensible and total cooling rates at a fixed indoor dry-bulb temperature under varying indoor wet-bulb temperatures and outdoor dry-bulb temperatures. The forecasted wet-bulb temperatures and the linear heat exchanger regression in Eq. \eqref{SHR} gave a SHR prediction RMSE of 0.1 and $R^2$ of 0.95. Fig. \ref{SHR_WEEK} shows the performance of the model in a typical week.

\subsection{Supervisory controller design}
\label{Scheme}

\textit{Formulation.} At each time step, the supervisory control system gathers the most recent temperature and power measurements, obtains the latest weather forecast, predicts trajectories of the exogenous thermal power and SHR, solves an open-loop optimal control problem to plan a trajectory of indoor temperature set-points, and sends the first planned set-point to the device-level control system. The process is shown in Fig. \ref{summer_flowchart}.  This subsection describes the open-loop optimal control problem's decision variables, objectives, constraints, and input data. The full formulation is shown in Eq. \eqref{eq:full_formulation}.

\begin{figure*}[!t]
\begin{subequations}
\label{eq:full_formulation}
\begin{align}
&\text{Power, energy, and comfort objectives:} && \min \limits_{T,\dot Q_{c},P} \: \pi_d \max_k ( P_k ) + \Delta t \sum\limits_{k=0}^{L-1} \left( \pi_e P_{k} + \pi_t|T_{\text{pref},k}-T_{k}|\right) \label{eq:objective} \\
&\text{Temperature dynamics:} &&\begin{aligned}[t] & T_{k+1} = \alpha T_{k} + (1 - \alpha) \left[ T_{\text{eq},k} + R \left( \dot Q_{c,k} + \dot Q_{e,k} \right) \right] \\
& T_{0} = T_{\text{initial, building}} \end{aligned} \label{eq:state_air} \\
&\text{Electrical power:} && P_{k} = \frac{\dot Q_{c,k}}{\text{SHR}_{k} \text{COP}_{k}} 
\label{eq:expressions} \\
&\text{Capacity and comfort constraints:} &&\begin{aligned}[t]
& 0 \leq P_k \leq P_{\text{HP,max}} \\
& 0 \leq \dot Q_{c,k} \leq  P_{\text{HP,max}} \text{COP}_k \\
& |T_{\text{pref},k} - T_k| \leq \delta  \label{eq:inequality_constraints} 
\end{aligned} 
\end{align}
\end{subequations}
\end{figure*}

The control objectives are to reduce the peak power, reduce the electrical energy used by the heat pump, and track a the occupants' thermal preference temperature, $T_\text{pref}$ (Eq. \eqref{eq:objective}). The discrete-time temperature dynamics are shown in \eqref{eq:state_air}. Eq. \eqref{eq:expressions} shows the mapping from the sensible heat supply to the electrical power, accounting for the total building load. Constraints were placed on the total allowable electrical power, sensible cooling rate, and indoor air temperature. 

The objectives and constraints in Eq. \eqref{eq:full_formulation} comprise the cost-reducing MPC optimization problem. The power-limiting MPC control problem includes an additional term in the objective function:
\begin{equation}
  \Delta t \pi_\text{peak} \sum\limits_{k=0}^{L-1} \max( P_{k} - P_{\text{lim},k}, 0)  .
\end{equation}
The power limit $P_{\text{lim},k}$ is assumed to be provided by the utility. The power limit is imposed only between 4 PM and 8 PM:
\begin{equation}
P_{\text{lim},k} = \begin{cases}
2.5 \text{ kW} & \text{if hour} \in [4, 8) \text{ PM} \\
\infty & \text{else}.
\end{cases}
\label{powerilim}
\end{equation}
This time period corresponds to system-wide electricity demand peaks in Indiana, as per the local electricity provider Duke Energy's Time-Of-Use rate plans. The price $\pi_\text{peak}$ was manually tuned to 1.4 \$/kWh, {\it i.e.} ten times the base electricity rate.

\begin{figure*}
\centering
\includegraphics[width=0.45\textwidth]{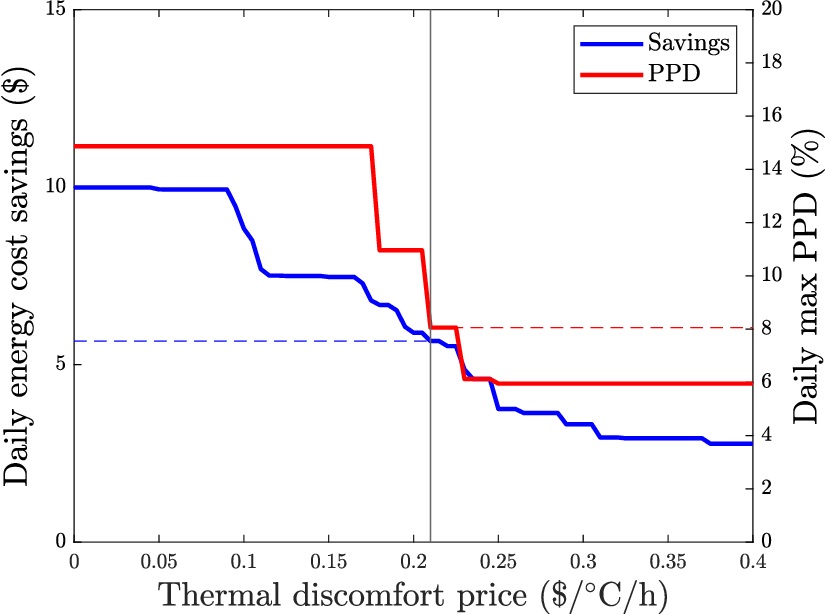}
\quad
\includegraphics[width=0.45\textwidth]{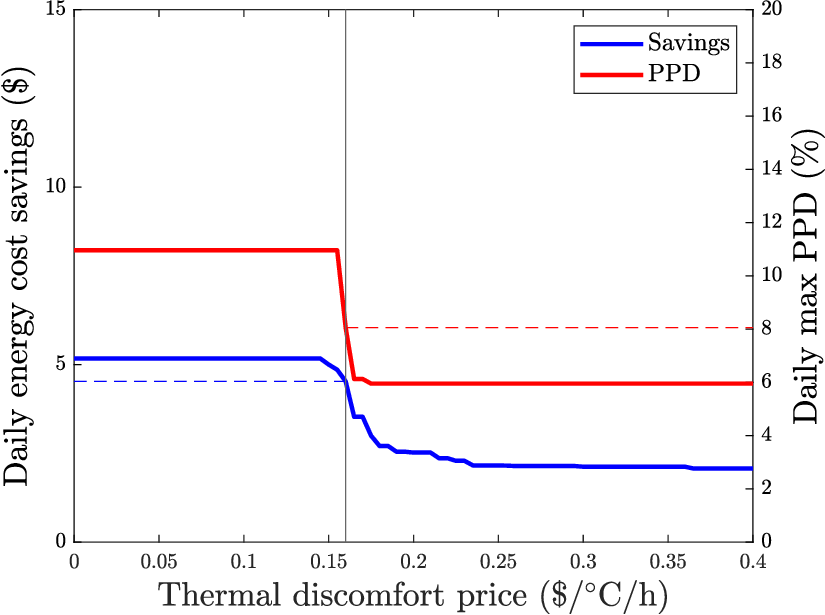}
\caption{The tuned thermal discomfort price (vertical line) was higher on a hot day (left, average outdoor temperature $28.8$ $^\circ$C) than on a mild day (right, 22.6 $^\circ$C). On both days, the simulated maximum PPD was less than 10\%. Savings are estimated using the thermal circuit model and the house's historical average indoor air temperature set-point of 23 $^\circ$C.}
\label{discomfortPriceTuning}
\end{figure*}

\textit{Optimization inputs.}
In field demonstrations of the supervisory control system, a time step duration of $\Delta t = 1$ h was used, prediction horizon of $L = 24$ time steps, constant electrical energy price $\pi_e = 0.14$ \$/kWh, heat pump capacity $P_\text{HP,max} = 4.5$ kW, temperature deviation $\delta = 3$ $^\circ$C, and trained discrete-time dynamics parameter $\alpha = 0.86$. The temperature deviation band was calibrated in simulation to ensure acceptable indoor conditions and PPD across varying ambient temperatures and humidities. The thermal resistance was $R = 1.04$ $^\circ$C/kW, with the outdoor conditions prediction obtained from weather forecasts. The COP is predicted in the case of the sensible formulation by propagating the outdoor temperature forecast through a quadratic function fit to manufacturer data, while the latent formulation uses both the outdoor temperature forecast and the GPR prediction of the indoor wet-bulb temperature. The exogenous thermal power $\dot Q_e$ is predicted by propagating weather predictions and time features through the support vector machine described in \cite{pergantis2024field}. The SHR is assumed to be a constant value of 0.86 for all time steps for the sensible model formulation and is predicted in the latent model formulation using the linear model in Fig. \ref{SHRFig}. The occupants specified preference temperatures $T_\text{pref}$ of 23 $^\circ$C, constant during the day. A daytime setback was not considered since the site has frequent visitors. A flat electricity price was used since that is the utility structure in place in West Lafayette, Indiana, where the test house resides. However, the control framework can accommodate any dynamic pricing scheme, such as time-of-use or real-time pricing. The optimization problem presented in Eq. \eqref{eq:full_formulation} is convex and solved in Matlab using the CVX toolbox \cite{cvx}. Infeasibilities rarely occur due to the soft treatment of the temperature bounds, but when they do, the control system defaults to the indoor temperature set-point from the previous time step. 

\textit{Price tuning.}
The formulation presented in Eq. \eqref{eq:objective} involves prices that govern trade-offs between multiple objectives, such as peak electricity demand, energy costs, and thermal comfort. Achieving good controller performance requires tuning the peak demand price $\pi_d$ and thermal discomfort price $\pi_t$. The utility serving the test-site does not impose a peak demand charge. Rather, the demand price is included as an additional incentive to mitigate demand peaks or provide power-limiting control. Tuning the peak demand price $\pi_d$ to \$0.8 per kW of daily peak demand, corresponding to a monthly peak demand price of about 25 \$/kW (a typical value for monthly peak demand charges in commercial buildings in the USA) gave good performance. 

Given $\pi_d$, the tuning problem reduces to calibrating the discomfort price $\pi_t$. Tuning $\pi_t$ interpolates between two extremes: for $\pi_t = 0$, there is no set-point tracking objective; while in the limit $\pi_t \rightarrow \infty$, the set-point is constrained to equal the user's preference $T_\text{pref}$. The supervisory control system automatically tunes the thermal discomfort price every 12 hours by sweeping an array of $\pi_t$ values, solving an open-loop optimal control problem for each value, and selecting the lowest $\pi_t$ that maintains the time-average PPD over the prediction horizon below 10\%. Once the sequence of set-points is obtained, the forecasted open-loop humidity and set-point time-series is fed through the full set of nonlinear PPD models as per the following equation:
\begin{equation}
    \pi_t = \min \{ \pi \mid \text{PPD}(T,T_{wb}) \text{ at price } \pi \leq 10\% \} . 
    \label{price_tuning_equation}
\end{equation}
The resulting value $\pi_t$ is conservatively increased by 10\% to account for any potential plant-model mismatch. This tuning procedure led to higher discomfort prices on warmer days, as Fig. \ref{discomfortPriceTuning} illustrates.

\begin{figure*}
\centering
\includegraphics[width=0.45\textwidth]{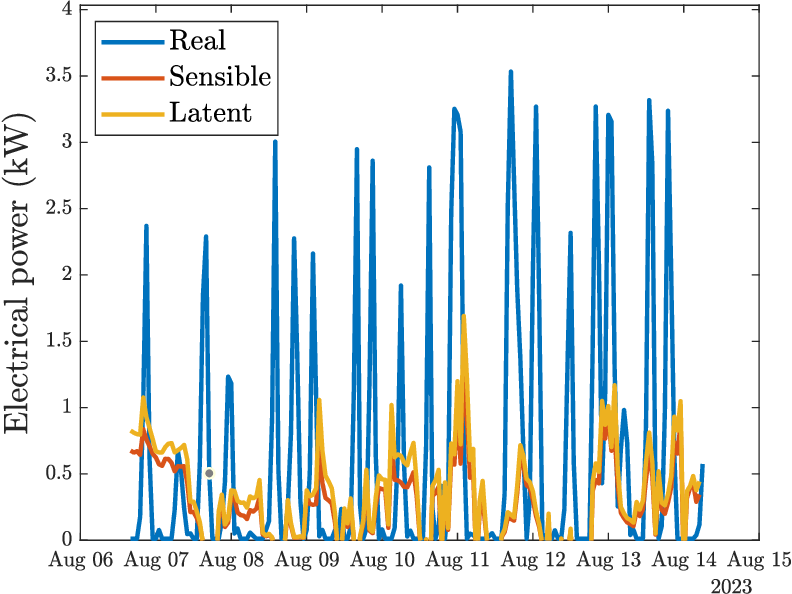}
\quad
\includegraphics[width=0.425\textwidth]{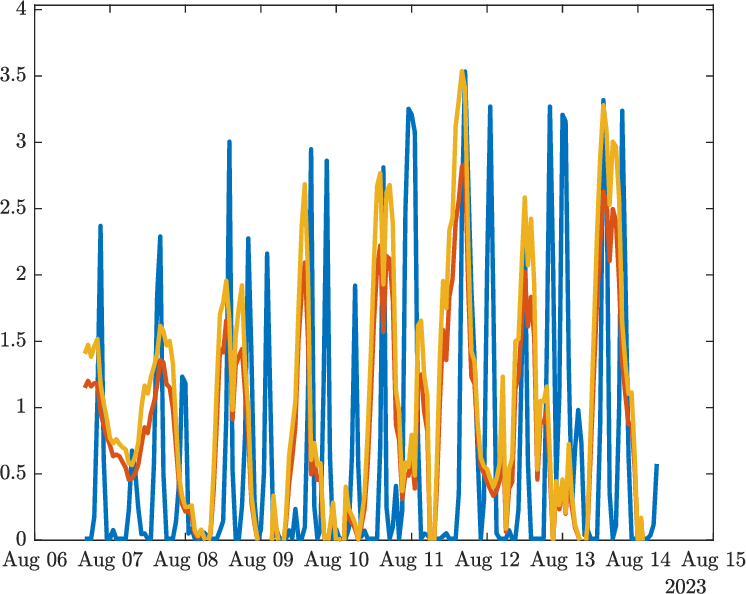}
\caption{Comparison of the overfitted thermal circuit model (left) with the recalibrated model (right). In both the sensible (orange) and latent (yellow) formulations, the model predicted peaks in measured electrical power (blue) much better after recalibration. However, the prediction RMSE decreased from 1.2 to 1.4 kW, suggesting that RMSE may not be the best training metric.}
\label{old_vs_new}
\end{figure*}

\subsection{Thermal circuit model re-calibration}
\label{new_rc}
Upon the completion of the cost-reducing MPC testing (24 days), it was found that although the controller can provide energy savings, the building model (Fig. \ref{modelFitFigure}) was unable to capture the true electrical peaks. This is shown across a typical week of testing in Fig. \ref{old_vs_new} where the model predictions are compared to the real values from the site. However, despite this, the controller was able to achieve energy savings (Sections \ref{energysavings}, \ref{powerlim}, and Fig. \ref{CDD_summer}). This corroborates the findings of other researchers \cite{blum2019practical, wang2023field} that as long the controller can reasonably predict the temperature at the next time step and the trajectory of the load, predicting the magnitude of the load with high accuracy is not necessary. This may not apply in cases that require very accurate power predictions, such as minimizing peak demand, constraining power below a specific target, or controlling thermally massive systems \cite{PICARD2017739}.

The right plot in Fig. \ref{old_vs_new} shows an improved thermal circuit fit, which involved manual tuning of the physical parameters. For the improved fit, $\alpha = 0.77$, $R = 0.42$ $^\circ$C/kW, and $\dot Q_e$ =  3.4 kW (constant) were used. The resulting model has an improved prediction of the one-step-ahead temperature (RMSE of 0.4 $^\circ$C), and while the prediction of the cooling rate is slightly worse (RMSE of 4.2 kW), it does not attempt to fit the mean of the data as the previous model.

\section{Building, hardware, and software}
\label{systemSection}

\subsection{Test House} 
This section briefly introduces the DC Nanogrid House (Fig. 1) as an experimental testbed; Pergantis et al. \cite{pergantis2024field} described the house and instrumentation in more detail. The DC Nanogrid House is a 208 m$^2$, two-story, 1920s-era detached single-family home near Purdue University's campus in West Lafayette, USA. This location falls under the International Energy Conservation Climate Code Zone 5A. This climate sees both hot summers (up to 35 $^\circ$C) and cold winters (down to $-20$ $^\circ$C). Climate Zone 5A is also humid. At the test site, for example, over 25\% of the experiment days had a dewpoint temperature greating than 20 $^\circ$C, indicating humid conditions \cite{Berglund1998}.

The DC Nanogrid House's exterior walls have foam insulation with an R-Value of 3.5 $^\circ$C m$^2$/W. Code-minimum U-8 W/m$^2$/$^\circ$C windows make up about 20\% of the exterior wall area. Networked sensors throughout the thermal and electrical systems have been installed. The house is a living laboratory occupied by graduate students.

\subsection{Cooling equipment} A central air-to-air heat pump heats and cools the DC Nanogrid House. The heat pump has 14 kW of rated cooling capacity, a cooling and heating seasonal COPs of 5.3 and 2.5, respectively. There is no mechanical intake of outdoor air, as fresh air enters via passive infiltration through the building envelope. The heat pump's indoor and outdoor fans and its compressor have variable-speed drives. The supervisory control system developed in this work decides indoor temperature set-points and sends them to the heat pump's thermostat.

\subsection{Sensing, communication, and computing} The supervisory control system uses measurements from Internet-connected power and temperature sensors. An IoTaWatt electric power meter monitors subcircuits and communicates via Wi-Fi to an InfluxDB database hosted by DigitalOcean. The database also stores the heat pump thermostat's indoor temperature measurement, extracted via Wi-Fi. A desktop computer, located at the DC Nanogrid House, periodically pulls measurements from the database, downloads day-ahead hourly weather forecasts from Oikolab, computes the next indoor temperature set-point, and pushes the set-point to the thermostat. More information on the setup can be found in \cite{Pergantis2023sensors}. Two Vaisala HMD65 sensors were installed to monitor the humidity in the return and supply air ducts. These sensors are pre-calibrated and accurate within $\pm$2.5\% of the measured relative humidity.

\section{Field demonstration results}
\label{resultsSection}

\subsection{Controller behavior} 
\label{modelaccuracy_section}
\begin{figure}
\includegraphics[width=0.45\textwidth]{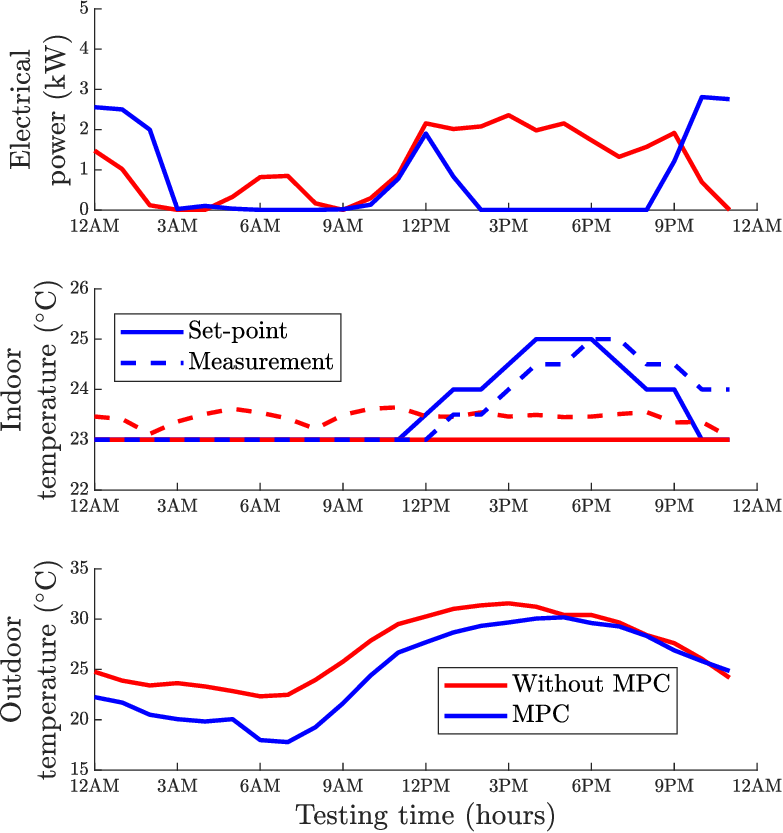}
\caption{On two similar days, the cooling equipment used 32\% less energy with MPC (blue, 17.7 kWh) than without MPC (red, 26.0 kWh) by shifting load from the hot afternoon to the cooler night (higher COP). Mean ambient relative humidity on both days was approximately 70\%.}
\label{typicalday}
\end{figure}

The different control system implementations were tested continuously from July 10th to September 7th 2023. Of these 60 days, seven were lost due to intermittent thermostat Wi-Fi issues, seven were lost due to controller issues (badly tuned comfort price, communication issues), and eight were reverted back to non-MPC control. Of the remaining 38 days, 24 used cost-reducing control, while 14 performed power-limiting control. For the power-limiting MPC, the re-calibrated thermal circuit model in Section \ref{new_rc} was used, while for the cost-reducing MPC, the original values presented in Section \ref{Scheme} were used. Across the 38 testing days, the sensible and latent controller formulations were tested in an one-day-on, one-day-off fashion, with a total of 21 latent days and 17 sensible. Fig. \ref{typicalday} shows the performance of the controller on a typical cost-reducing MPC day. The set-point typically increases during the hot afternoon, when the COP is low, to save electricity. The set-point typically returns to the occupants' preference overnight and through the morning.

\subsection{Thermal comfort} The supervisory control system's thermal comfort performance was assessed in two ways. First, a web form was offered for occupants to record any thermal discomfort they felt. During the total duration of testing, only one response was received, and that was of mild dissatisfaction (as per \cite{enescu2017review}). Second, a thermal comfort software package \cite{tartarini2020pythermalcomfort} was used to generate a PPD time series based on indoor temperature measurements and other inputs to the PPD model, such as occupants' estimated clothing and activity levels. This process yielded a time-average PPD of 7.2\% over the MPC test hours and 8.5\% over the test hours without MPC. Practitioners typically consider $\leq 10$\% PPD acceptable \cite{enescu2017review}. The PPD was in the slightly uncomfortable region for a total of two hours across the MPC testing days. While these PPD results are encouraging, they are not definitive. PPD may not accurately reflect comfort under highly transient conditions \cite{CHENG201213}. Furthermore, the PPD calculations here assumed constant air flows, activity levels, and clothing levels; in reality, these varied somewhat over the experiments. Nevertheless, the combined results of the PPD calculations and occupant questionnaires suggest that MPC maintained satisfactory comfort with both the latent and sensible MPC models.

\begin{figure}
\includegraphics[width=0.45\textwidth]{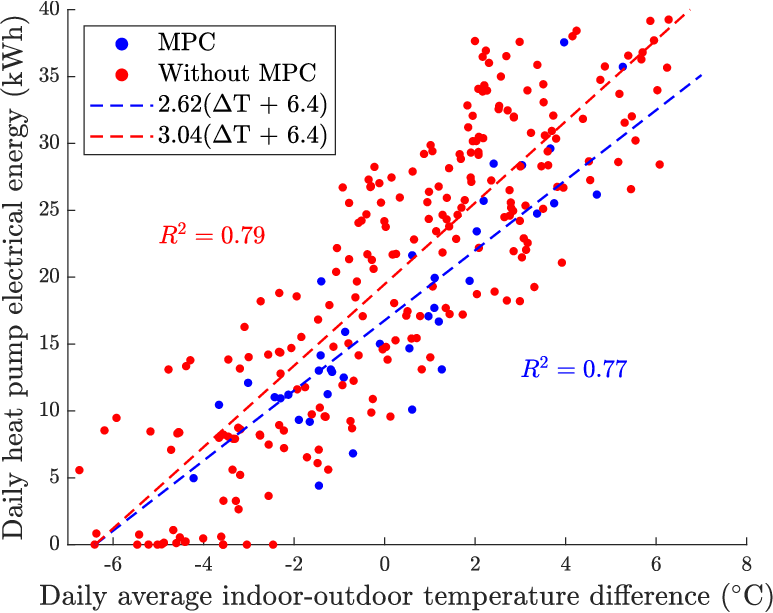}
\caption{Energy use for both energy-reducing and power-limiting MPC (blue, mixed mode). The supervisory controller typically saves about 14\% of daily cooling
electrical energy. Absolute savings increase with the difference between the indoor and outdoor air temperatures.}
\label{CDD_summer}
\end{figure}

\subsection{Energy savings} 
\label{energysavings}
\textit{Differences across controller modes.}
Fig. \ref{CDD_summer} shows the daily electrical cooling energy with MPC (blue) across all formulations and without MPC (red), as a function of the daily average temperature difference between the indoor and outdoor air, $\Delta T$ ($^\circ$C). Energy use increases approximately linearly with the temperature difference both with and without MPC, but the slope is higher without MPC. Under the linear fits (dashed lines) in Fig. \ref{CDD_summer}, the slopes are approximately Gaussian distributed. With MPC, the mean and standard deviation of the slope $m_1$ are 2.62 and 0.096 kWh/$^\circ$C. Without MPC, the mean and standard deviation of the slope $m_2$ are 3.04 and 0.049 kWh/$^\circ$C. The relative daily energy savings, as a fraction of the non-MPC daily energy use, are given by: \begin{equation}
\begin{aligned}
1 - \frac{\text{MPC energy}}{\text{Non-MPC energy}} &\approx 1 - \frac{m_1 (\Delta T + 6.2) }{ m_2 (\Delta T + 6.2) }\\
&= 1 - m_1 / m_2 .
\label{slope_results}
\end{aligned}
\end{equation}

Using Eq. \eqref{slope_results}, a Monte Carlo simulation is performed by generating 10$^7$ samples from the slope distributions for MPC and non-MPC to obtain the distribution of the savings \cite{pergantis2024field}. The energy savings across the 38 days of testing (both cost-reducing and power-limiting MPC) are 7-21\% (95 \% confidence interval) with a sample mean of 14\%.

\begin{figure}
\includegraphics[width=0.45\textwidth]{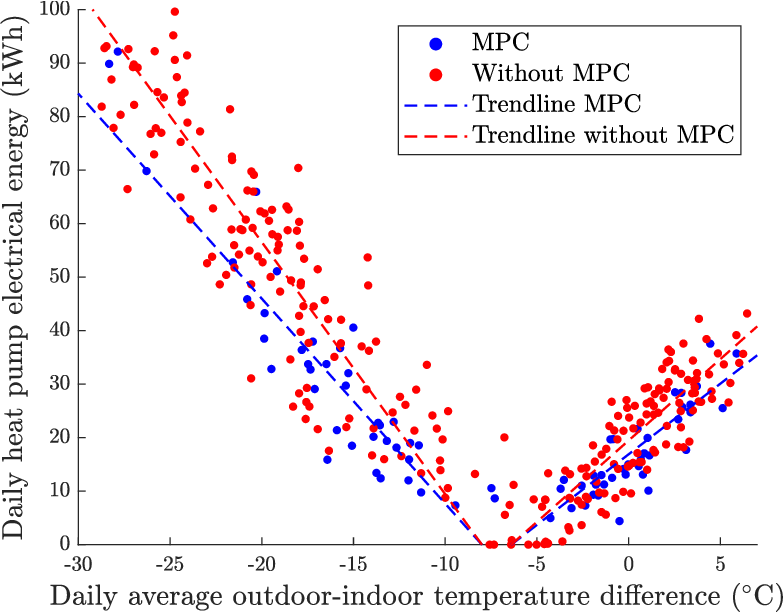}
\caption{In 71 days of winter and summer testing, MPC (blue) significantly improved energy efficiency.}
\label{HDDCDD}
\end{figure}

Subsequent analysis broken down by controller formulation (cost-reducing vs. power-limiting) found that, as expected, the cost-reducing MPC performed better (16.6 to 32.5\% energy savings). However, the power-limiting MPC also offered some moderate energy savings (-5 to 10\%). These differences were attributed to two factors. First, there were differences in outdoor temperatures across the test days for the two formulations. Power-limiting MPC was tested on hot days in mid-late August that were on average 2 $^\circ$C hotter than for cost-reducing control. Second, power-limiting MPC typically pre-cooled the building during the hot afternoon, when the COP was typically low, in anticipation of the 4 to 8 PM demand response window. Zhang et al. \cite{zhang2022model} observed similar behavior from a power-limiting controller in a commercial building. 

\begin{figure}
\includegraphics[width=0.45\textwidth]{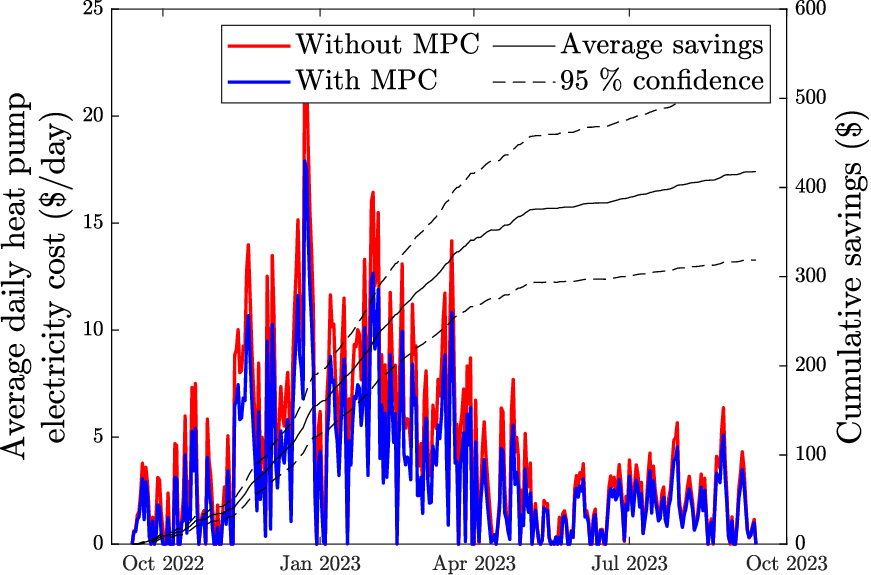}
\caption{Daily heat pump operation cost (left axis) with MPC (blue) and without (red) as well as the resulting cumulative cost savings (right axis). The dashed lines indicate the 95\% confidence interval.}
\label{savings_timeseries}
\end{figure}

\begin{figure*}
\centering
\includegraphics[width=0.48\textwidth]{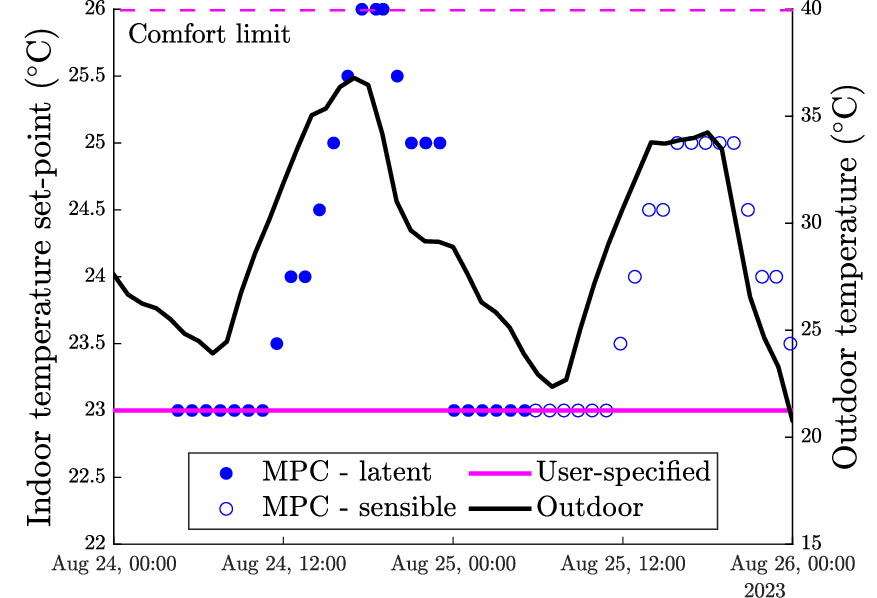}
\quad
\includegraphics[width=0.45\textwidth]{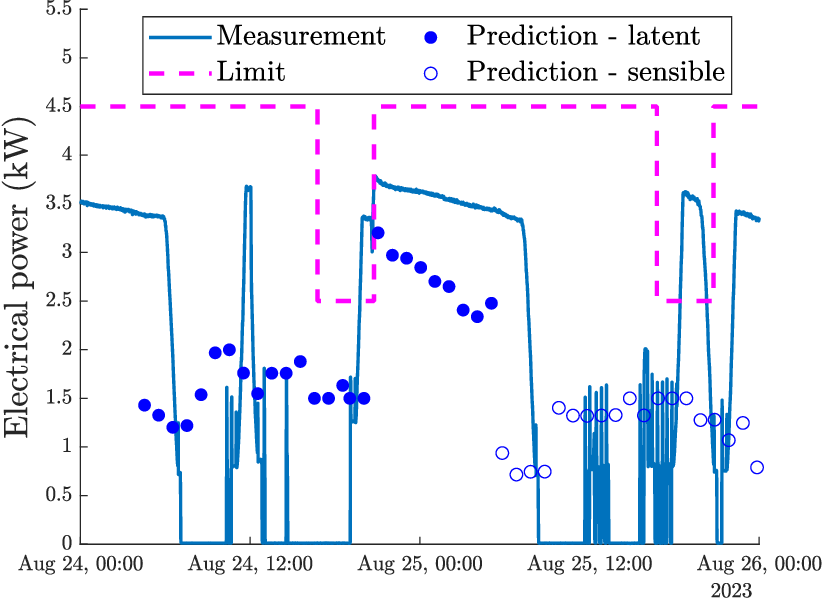}
\caption{Performance of power-limiting control over two warm days for temperature (left) and electrical power (right). While the sensible MPC (empty circles) often violates the power constraint, the latent MPC satisfies it most of the time.}

\label{plim}

\end{figure*}

\textit{Yearly cost-reducing performance.} Fig. \ref{HDDCDD} shows MPC performance over 71 total days of testing, including summer testing results from this work and winter testing results from Pergantis et al. \cite{pergantis2024field}. Fig. \ref{HDDCDD} shows that significant energy savings are achievable with MPC for both cooling and heating. The temperature-normalized energy models in
Fig. \ref{HDDCDD} enable estimation of the total electricity cost savings from MPC in the DC Nanogrid House over an entire year, from September 13th of 2022 to 2023. The estimation methodology is identical to the one presented for heating in Pergantis et al. \cite{pergantis2024field}, extended to incorporate the cooling season. The daily expected costs associated with different methods, as well as the cumulative savings, are shown in Fig. \ref{savings_timeseries}. The annual savings are found to be between \$340 to 497 (95\% confidence interval), with a sample mean of \$419. The estimated annual percentage cost savings were 27\% (22\% to 31\%).

\begin{figure*}[ht!]
\centering
\includegraphics[width=0.45\textwidth]{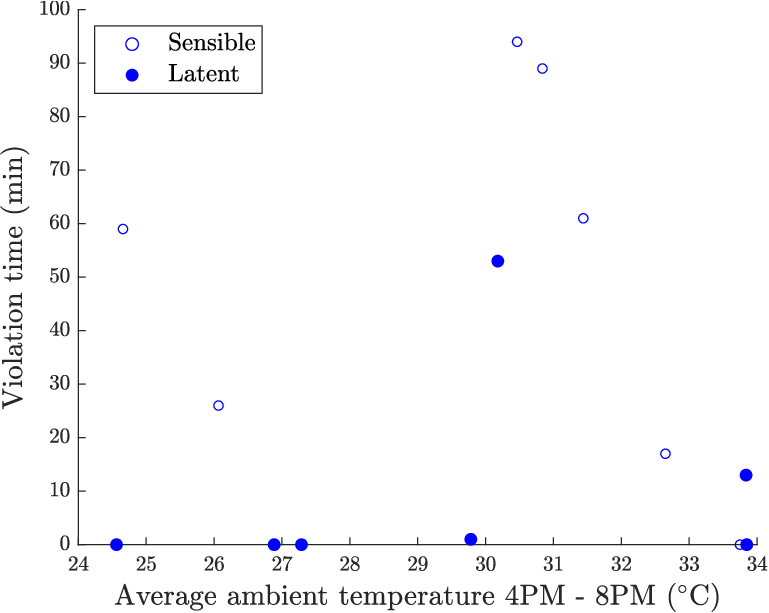}
\quad
\includegraphics[width=0.45\textwidth]{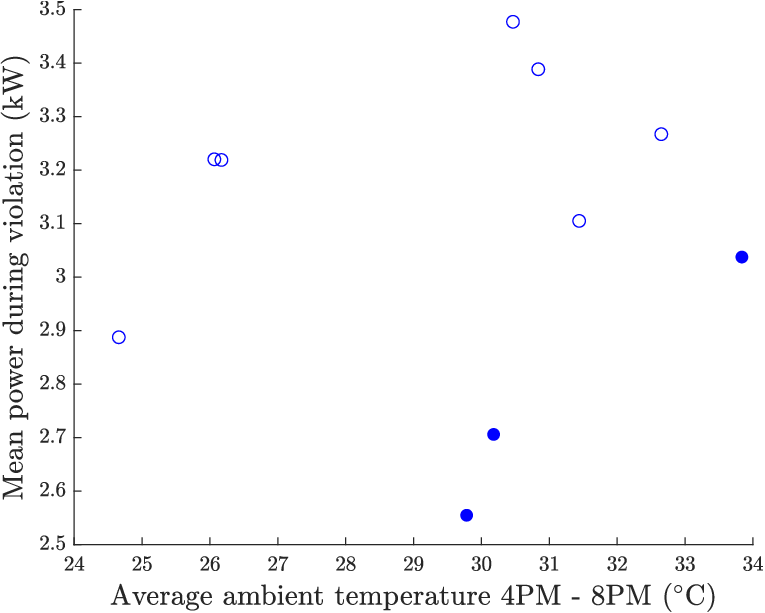}
\caption{Comparison of the violation time (left) and average magnitude if the violation occurred (right) for the two model structures. The latent model violates the constraint less often and when it does, for a shorter duration and lower magnitude.}
\label{powerlimiting_performance}
\end{figure*}

\subsection{Power limiting} 
\label{powerlim}

\begin{table*}[bt]
\centering
\caption{Comparison of different testing modes and model types.}
\label{table:controller_comparison}
\resizebox{0.9\textwidth}{!}{%
\begin{tabular}{c|cccc}
\begin{tabular}[c]{@{}c@{}}Cost-reducing\\ MPC\end{tabular} &
  \begin{tabular}[c]{@{}c@{}}Testing duration\\ (days)\end{tabular} &
  \begin{tabular}[c] {@{}c@{}}Weather-normalized energy \\  (kWh/$^\circ$C) \end{tabular} &
  \begin{tabular}[c]{@{}c@{}}Mean constraint\\ violation duration (min/day) \end{tabular} &
  \begin{tabular}[c]{@{}c@{}}Conditional-mean constraint\\ violation magnitude (kW)\end{tabular} \\ \hline
Sensible                                                                                  & 10 & 2.32 & -   & -  \\
Latent                                                                                    & 14 & 2.34 & -   & -  \\ \hline
\multicolumn{1}{c|}{\begin{tabular}[c]{@{}c@{}}Power-limiting\\ MPC\end{tabular}} &    &      &     &    \\ \hline
Sensible                                                                                 & 7  & 2.96 & 54 & 0.8 \\
Latent                                                                                    & 7  & 3.01 & 11 (80\% reduction) & 0.3 (63\% reduction) \\ \hline
\end{tabular}%
}
\end{table*}

After testing with the cost-reducing controller, the system was switched to power-limiting MPC.  The key performance indicators under power-limiting MPC included (a) the mean number of minutes that the power surpassed the specified power limit (Eq. \eqref{powerilim}), and (b) the conditional-mean magnitude of constraint violation, conditioned on the event that the constraint was violated.
 
The results of testing over a hot week are shown in Fig. \ref{plim}. The latent formulation predicted the true electrical peaks more accurately, allowing it to drift the required amount to minimize the power use over the window. Fig. \ref{powerlimiting_performance} shows the performance of the MPC algorithms in mitigating power peaks. MPC with the latent formulation reduces both the the magnitude and frequency of constraint violation. sensible MPC violated the power limit between 4 and 8 PM for an average of 54 minutes per day, with a conditional-mean magnitude of 0.8 kW. By contrast, latent MPC violated the constraint for an average of 11 minutes per day, with a conditional-mean magnitude of 0.3 kW. Relative to sensible MPC, latent MPC reduced the daily mean duration of constraint violation by 80\% (from 54 to 11 minutes) and the conditional-mean magnitude of constraint violation by 63\% (from 0.8 to 0.3 kW).

\subsection{Comparison of humidity modeling approaches} 

Table \ref{table:controller_comparison} compares the performance of MPC with the sensible and latent humidity modeling approaches, for both energy efficiency and power limiting. In this table, the weather-normalized energy is defined as the ratio of the cumulative daily heat pump electrical power to the cumulative daily indoor-outdoor temperature difference, averaged over $N$ total testing days:
\begin{equation}
    \frac{\Delta t \sum_{n = 1}^{N} \sum_{k = 1}^{24} P_{k,n} }{\sum_{n = 1}^{N} \Delta T_{n}} .
\end{equation}
A high weather-normalized energy indicates higher energy consumption for the same indoor-outdoor temperature difference. For comparison, the weather-normalized energy was 3.14 kWh/$^\circ$C under benchmark control without MPC. 

Table \ref{table:controller_comparison}  shows that sensible and latent MPC performed similarly for energy efficiency, with weather-normalized energies of 2.32 and 2.34 kWh/$^\circ$C, respectively, for cost-reducing MPC.  However, latent MPC significantly outperformed sensible MPC for power-limiting control, reducing the mean duration of constraint violation by 80\% and the conditional-mean violation magnitude by 63\%. These results suggest that model accuracy matters more when objectives or constraints depend sensitively on the predicted magnitude of power use. 

These results, alongside prior literature \cite{RAMAN2020115765, RAMANmanyclimates, AGHNIAEY201819}, suggest that accurate humidity modeling can improve performance from some residential HVAC control tasks. For other tasks, however, good performance can be achieved without modeling humidity accurately, or indeed modeling it at all \cite{KIM201849, HAM2023113351}. For example, Table \ref{table:controller_comparison} shows that despite its oversimplified humidity model, sensible MPC still reduced the mean duration of violating the power-limiting constraint by 40\% relative to the non-MPC baseline, from 90 minutes of violation to 54.

\section{Discussion of results and challenges}
\label{discussionSection}
\subsection{Practical considerations}

\textit{Scalability.} To reduce energy bills, greenhouse gas emissions, and power-grid impacts at scale in the real world, a supervisory control system for residential air conditioning should be inexpensive and easy to deploy. For the thermal envelope training, the field testing in this study used precise lab-scale sensors to estimate the air conditioner's real-time cooling rate from temperature, humidity, and flow rate measurements of the indoor air stream. These sensors are expensive and require expertise to install and commission. In practice, the cooling rate could instead be estimated from a COP curve and electrical power measurements obtained from sensors that are less expensive and easier to install.

\begin{figure}
\includegraphics[width=0.45\textwidth]{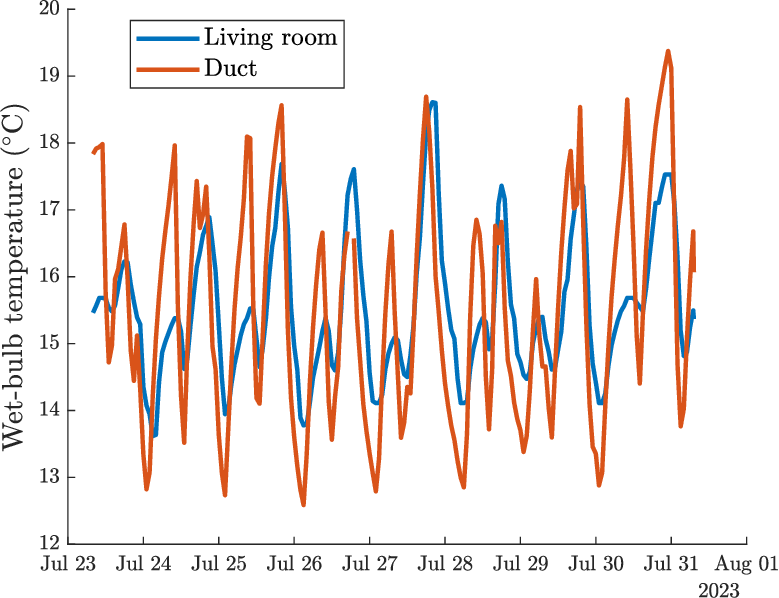}
\caption{Comparison of the wet-bulb temperature calculated using measurements at the thermostat (blue) and in the duct (orange). Thermostat measurements may fail to capture wet-bulb peaks, which could pose challenges for accurately predicting latent loads and electricity demand peaks.}
\label{duct_livingroom}
\end{figure}

The field tests in this study also used wet-bulb temperature measurements in the return air stream to train a predictive wet-bulb temperature model. HVAC manufacturers typically do not include return air wet-bulb temperature sensors in residential equipment, and installing such a sensor is likely to be prohibitively expensive in practice. To overcome this challenge, duct humidity conditions could be estimated from the indoor air humidity and temperature, which most thermostats measure. This approach would need to be robust to differences between the air conditions at the thermostat and in the return air duct, as shown in Fig. \ref{duct_livingroom}.

\textit{Impact of assumptions.}
One assumption underlying this work was the numerical value of the SHR used for the sensible MPC formulation. After testing, it was found that the conditions were more humid than expected: the SHR actually averaged 0.79, while the sensible MPC assumed an SHR of 0.86. It is challenging to select an appropriate SHR value, as (1) it can change from house to house based on floor space and envelope characteristics, (2) it is highly dependent on the weather and on internal moisture generation driven by occupant behavior, and (3) it can vary based on the capacity of the cooling equipment and heat exchanger characteristics. Typically, to avoid these issues, researchers do not model the latent load or the indoor humidity in experiments of supervisory control \cite{bunning2020experimental, KIM201849, wang2023field}. Another limitation of this study lies in the testing of the controller in a single climatic zone and a single household. More data are needed to assess performance under varying occupant preferences and housing types across different climates.

\subsection{Deployment challenges} 
Pergantis et al. \cite{pergantis2024field}, as well as Blum et al. \cite{blum2022field}, discuss the practical challenges of supervisory HVAC control. This study represents the second time that a supervisory HVAC control system was demonstrated at the DC Nanogrid House. It was found that, once stable data acquisition and communication systems were in place, efforts to develop the controller were significantly lower. However, some problems were still faced.

\textit{Air-to-air centrally ducted systems.} Most residential buildings in the USA have central ducted air conditioning systems. This equipment configuration presented two challenges during our testing. First, during winter testing, the return air temperature was used as the mixed house temperature. Due to stratification effects, this is conservative during winter, since the air going into the duct from near the floor of the rooms is typically slightly colder. However, during summer the opposite is true: If an in-duct sensor is used, the house might appear cooler than it is. This issue could be overcome by using the thermostat's temperature measurement, or by estimating the return air temperature using a Kalman filter. Second, when quantifying comfort, thermal zoning should be accounted for. In many houses, poor balancing of the return and supply air across different floors and rooms, as well as drafts or thermal bridging through a subset of the walls and windows, can result in significantly different temperatures in different rooms. In our formulation, a safety factor of one $^\circ$C was used in the temperature fed to the PPD model.

\textit{Using thermostat measurements.} A proprietary thermostat from a popular heat pump manufacturer was used in this study since variable speed equipment often cannot operate with 3rd party thermostats. Typically, high-end thermostats are accurate within $\pm$0.5 $^\circ$C for temperature, and, $\pm$5 to 10\% for humidity. These sensors also have a smoothing effect, as they average over previous values. Using the thermostat temperature measurements, it was challenging to learn a thermal circuit model that did not overfit the training data or require manual tuning. This was the source of the issues in Section \ref{modelaccuracy_section} since the methodology developed in prior work for training a low-order thermal circuit model was highly dependent on utilizing overnight indoor temperature measurements \cite{pergantis2024field}. However, the temperature readings from the thermostat were very slow under natural convection, reading almost a constant overnight temperature when the air conditioning system was off. Other temperature sensors in the house drifted a couple of $^\circ$C during that period. 

\textit{Network connectivity.} Major disruption in testing occurred due to communication issues. This was attributed to the router presenting an issue with the 2.4 GHz network, at which frequency typically smart thermostats and sensors operate. This resulted in the smart thermostat going intermittently offline. The thermostat manufacturer did not provide relevant information on their cloud Application Programming Interface call, rather continuing to propagate the last reading posted online. This made offline periods hard to diagnose. A workaround to this issue would be to send all the set-points over the forecast horizon to be stored on the thermostat's memory, rather than a single value every hour, so good (if not optimal) actions can continue to be implemented while the connection is reestablished.

\section{Conclusion}
This paper presented the first supervisory control experiment for residential air conditioning that accounted for time-varying humidity effects. It developed novel data-driven time-series forecasting methods to predict indoor humidity conditions. This paper tested MPC with two humidity models: a sensible model that assumed a constant SHR, and a latent model that predicted a time-varying SHR based on dynamic predictions of the return air wet-bulb temperature. Each modeling approach was tested in two MPC formulations: one that sought to minimize energy costs, and one that sought to constrain electrical power below a utility-specified limit during a demand response event. sensible and latent MPC performed similarly for energy cost reduction, but latent MPC performed much better for power limiting. These results corroborate a finding of other researchers -- that energy cost savings depend weakly on model accuracy -- but suggest that model accuracy is more important when minimizing or constraining peak power demand. 

Overall, the control approach demonstrated in this paper and in \cite{pergantis2024field} performed well. MPC maintained or improved thermal comfort relative to the manufacturer's default controls in both summer and winter, while reducing annual HVAC energy costs by an estimated \$419 (27\%). However, the control approach's generalizability to other housing types and climate zones remains to be demonstrated. Future research could also focus on reducing the implementation costs, for example by eliminating the need to measure temperature, humidity, or flow in the return air duct.

\section*{CRediT authorship contribution statement}

{\bf Elias Pergantis:} Conceptualization, Methodology, Software, Investigation, Formal Analysis, Data Curation, Visualization, Writing – original draft, Writing – review \& editing. {\bf Parveen Dhillon:} Conceptualization, Methodology, Writing – original draft, Writing – review \& editing. {\bf Levi D. Reyes Premer:}  Software, Investigation, Writing – original draft, Writing – review \& editing. {\bf Alex H. Lee:}  Software, Investigation, Writing – original draft, Writing – review \& editing. {\bf Davide Zivani:} Methodology, Funding Acquisition, Writing - Review \& Editing, Project administration. {\bf Kevin Kircher:} Conceptualization, Methodology, Formal analysis, Writing - Review \& Editing, Visualization, Project administration.

\section*{Declaration of competing interest}

The authors declare that they have no known competing financial interests or personal relationships that could have appeared to influence the work reported in this paper.

\section*{Data availability}

Data will be made available on request.

\section*{Acknowledgments}

The Center for High-Performance Buildings (CHPB) at Purdue University supported this work (project CHPB-26-2023). E. Pergantis was also supported by the Onassis Foundation as one of its scholars, as well as the American Society of Heating and Refrigeration Engineers (ASHRAE) through a Grant-in-Aid award. The authors would like to thank the occupants of the DC Nanogrid House for their patience during testing.

\section*{Appendix. Acronyms and notation}
\label{notation}

This paper used eight acronyms: COP (coefficient of performance), DC (direct current), GPR (Gaussian process regression), HVAC (heating, ventilation and air conditioning), MPC (model predictive control), PPD (predicted percentage dissatisfied), RMSE (root-mean-square error), and SHR (sensible heat ratio). Table \ref{notationTable} summarizes the mathematical notation used in this paper.

\begin{table}[ht!]
\caption{Mathematical notation} 
\label{notationTable} 
\centering
\small

\begin{tabular}{ l | l }
Symbol (Units) & Meaning \\
\hline
$\alpha$ (-) & Discrete dynamics parameter \\
$a$ (1/$^\circ$C) & Linear SHR slope \\
$b$ (-) & Linear SHR offset \\
$\gamma$ (kWh/$^\circ$C) &  Normalized daily energy use \\
$h$ (-) & Hour of day (0-23) \\
$I_{\text{solar}}$ (W/m$^{2}$) & Global horiz. solar irradiation \\
$L$ (-) & Prediction horizon \\
$m_1$ (kWh/$^\circ$C) & MPC energy savings slope \\
$m_2$ (kWh/$^\circ$C) & Non-MPC energy savings slope \\
$P$ (kW) & Heat pump electric power \\
$P_{HP,max}$ (kW) & Heat pump capacity \\
$P_{\text{lim},k}$ (kW) & Demand response power limit \\
$R$ ($^\circ$C/kW) & Effective resistance \\
$R_\text{out}$ ($^\circ$C/kW) & Indoor-outdoor resistance \\
 
$R_m$ ($^\circ$C/kW) & Air-mass resistance \\
 $\delta$ ($^\circ$C)  & Maximum temperature deviation \\
$\Delta T$ ($^\circ$C)  & Daily  indoor-outdoor temp. difference \\
$\Delta t$ (h) & Time step duration \\
$k$ (-) & Discrete time indices \\
$\pi_d$ (\$/kW) & Peak demand price \\
$\pi_e$ (\$/kWh) & Electricity price \\
$\pi_\text{peak}$ (\$/kWh) & Power-limiting window price \\
$\pi_t$ (\$/$^\circ$Ch) & Thermal discomfort price \\
$\dot Q_c$ (kW) & Heat pump thermal power \\
$\dot Q_e$ (kW) & Exogenous thermal power \\
$t$ (h) & Time \\
$T$ ($^\circ$C) & Indoor air temperature \\
$T_{\text{out}}$ ($^\circ$C) & Outdoor air temperature \\
$T_{\text{eq}}$ ($^\circ$C) & Equivalent boundary temperature \\
$T_\text{pref}$ ($^\circ$C) & Preference indoor temperature \\
$T_{wb}$ ($^\circ$C) & Wet-bulb temperature \\
\end{tabular}
\end{table}

\bibliographystyle{elsarticle-num}
\bibliography{refs.bib}

\end{document}